\documentclass{article}

\usepackage{graphicx}
\usepackage{psfig}
\usepackage{epsfig}
\usepackage[round]{natbib}

\title{
Statistical modelling of tropical cyclone tracks: non-normal innovations
}

\begin{document}

\author{Tim Hall, GISS\footnote{\emph{Correspondence address}: Email: \texttt{tmh1@columbia.edu}}\\and\\
Stephen Jewson\\}

\maketitle

\begin{abstract}
%We describe results from the sixth stage of a project to build a
%statistical model for hurricane tracks. Previous models have considered the genesis,
%track shape and lysis of hurricanes. One assumption in the models for the track shape was that
%the residuals were normally distributed, although this is not correct. We now test
%to see if results can be improved by considering non-normal innovations.
We present results from the sixth stage of a project to build a statistical hurricane model.
Previous papers have described our modelling of the tracks, genesis, and lysis of hurricanes.
In our track model we have so far employed a normal distribution for the residuals when
computing innovations, even though we have demonstrated that their distribution is not normal.
Here, we test to see if the track model can be improved by including more realistic non-normal innovations.
The results are mixed. Some features of the model improve, but others slightly worsen.
\end{abstract}

\section{Introduction}

There is considerable interest in modelling the shapes and intensities of tropical cyclone tracks,
and, in particular, where those tracks make landfall.
For example,
models have been proposed by ~\citet{fujii}, \citet{drayton00}, \citet{vickery00} and \citet{emanuel05}.
The development of these models has mostly been motivated by the interests of the insurance industry,
although one would imagine that
the results from such models could also be used by others who are affected by tropical cyclones.

None of the models cited above is entirely satisfactory, and
given the increasing damage caused by tropical cyclones
it seems appropriate to revisit the question of how to build such models.
To this end, we have undertaken to build a new tropical cyclone track and intensity model from scratch,
paying particular attention to using well-defined and appropriate statistical methodologies.
Our focus is on the Atlantic, although the methods we are developing should be directly applicable to other basins.

We have made considerable progress towards building a new tropical cyclone simulation model.
Our previous papers have considered how to model the following components:
\begin{itemize}
 \item the mean tracks of hurricanes~\citep{hallj05a}
 \item the variance of fluctuations around the mean tracks~\citep{hallj05b}
 \item the autocorrelation of fluctuations around the mean tracks~\citep{hallj05c}
 \item the annual pattern of hurricane genesis~\citep{hallj05d}
 \item the annual pattern of lysis (death) of hurricanes~\citep{hallj05e}
\end{itemize}

The results shown in~\citet{hallj05e} show that, in combination, these five models
do a good job of simulating the main features of hurricane track
behaviour, but that they are also deficient in some respects when compared with the observations in detail.
In particular, the area immediately off the coast of Georgia and the Carolinas
does not show sufficient track density (see figure 6 of~\citet{hallj05e}).
In this paper we test whether this problem can be solved by more accurate modelling of the residuals.
Diagnostics in~\citet{hallj05c} showed that these residuals were not normally distributed, although
up to now we have modelled them as normal (i.e. we have forced the simulations with normal innovations).
We now attempt to model the distribution of the residuals more accurately
by saving the residuals from the fitting process,
and resampling them to create the innovations that force the simulation model.
This effectively guarantees
that the innovations have the correct distribution. We then test the model at a macroscopic level
to see whether this improves the overall track simulation.

\section{Method}

The model used in~\citet{hallj05e} uses normally distributed innovations (noise forcing)
to generate track simulations, and the macroscopic results are reasonably good, as shown
in that paper. However, a detailed analysis of the residuals from the model, as described
in~\citet{hallj05c}, has shown that, in fact, the residuals from the model are not normally distributed.
In particular, figure 11 of that paper shows that the residuals are close to normally distributed
up to around plus and minus two standard deviations, but that beyond that are fat-tailed i.e.
have higher probability density than the normal distribution. This suggests that simulations
forced by normally distributed random noise are never likely to be exactly correct. Fitting a normal
distribution to the long-tailed residuals is likely to give a distribution that has too high
probability density in the central range of the distribution, and too low probability density in
the tails. Exactly how such microscopic details of the model then affect the macroscopic behaviour of
the simulated tracks is hard to understand, but one might imagine that the resulting simulated tracks
might be too variable, but with too few really extreme fluctuations.

How should we deal with this situation? There are a number of possible approaches. First, it is possible that
we could reformulate the model in such a way that the residuals become normal.
For example, we could transform the fluctuations around the mean tracks to normal distributions,
which would likely lead to normal residuals.
Alternatively, we could try to avoid reformulating the model, and simply
accept the non-normality of the residuals and model it directly. After all, why should the residuals
be normally distributed in the first place? This latter approach is the approach that we take in this paper.
Since there is no obvious parametric distribution to use to model these fat-tailed residuals, we use
a simple non-parametric approach. The residuals from the fitting process are stored. During the simulations,
they are then resampled randomly and used as innovations.

There are a number of assumptions built into this method, that we haven't checked at this point.
In particular, we are assuming that the residuals are identically distributed throughout the basin,
throughout the season, and for storms of different origin.
In detail this assumption is likely to be wrong, and there are very likely to be
variations in the distributions of the residuals with all these three factors. The question is, whether these
variations are large enough to warrant making the simulation model more complex to incorporate them.
We will leave this question for a future study.

There is also a slight contradiction inherent in our model, in that the smoothing lengthscales for variance
and autocorrelation are fitted using the likelihood for a normal distribution, even though, as mentioned
above, the distribution is not exactly normal.

\section{Results}

We now show some results from applying the non-normal residual simulations described above.
These results should all be compared with the equivalent results for the previous version
of the model, described in~\citet{hallj05e}.

\begin{figure}[!hb]
  \begin{center}
    \scalebox{0.8}{\includegraphics{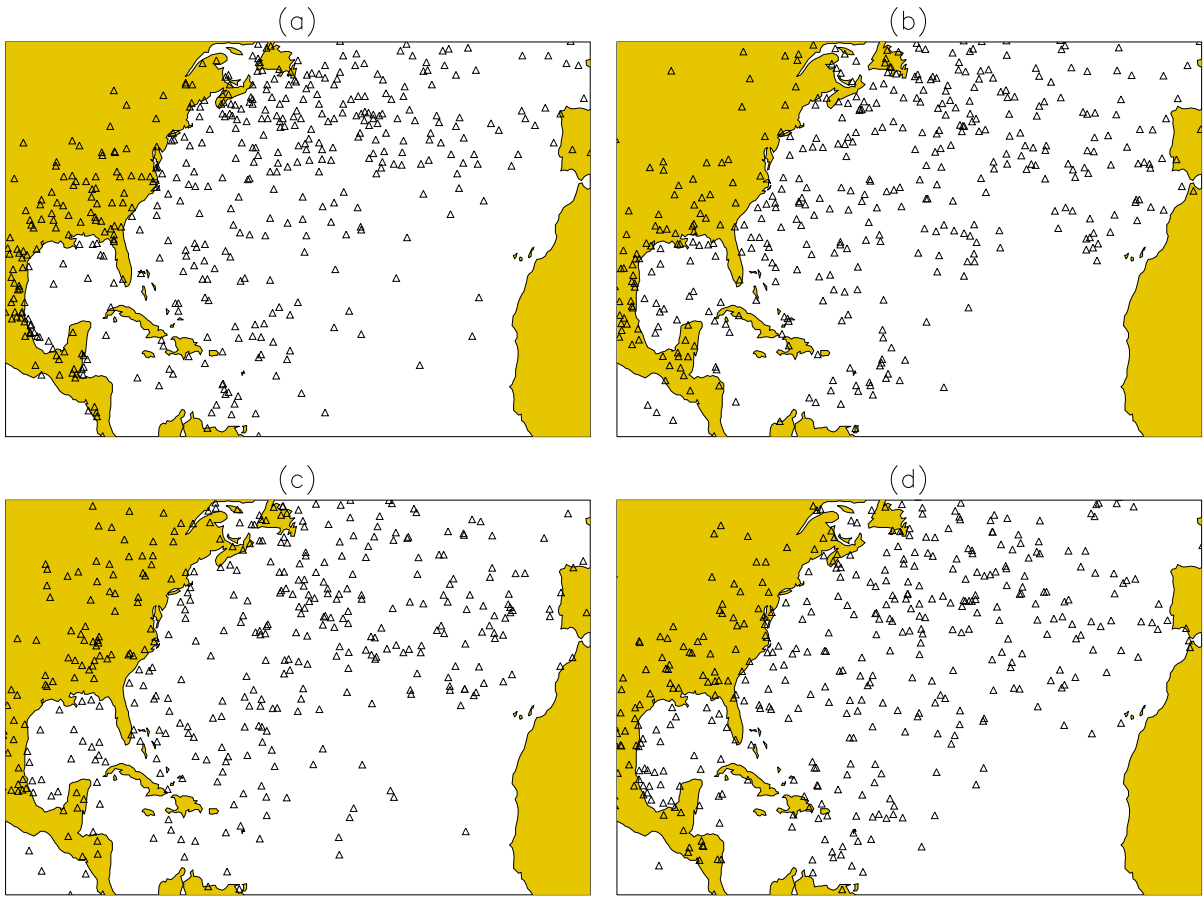}}
  \end{center}
    \caption{
The observed locations of hurricane lysis for the 524 storms from 1950 to 2003 (panel (a)),
and three simulations of hurricane lysis, each for 524 storms, for the model described in the text.
     }
     \label{f00}
\end{figure}

\begin{figure}[!hb]
  \begin{center}
    \scalebox{0.8}{\includegraphics{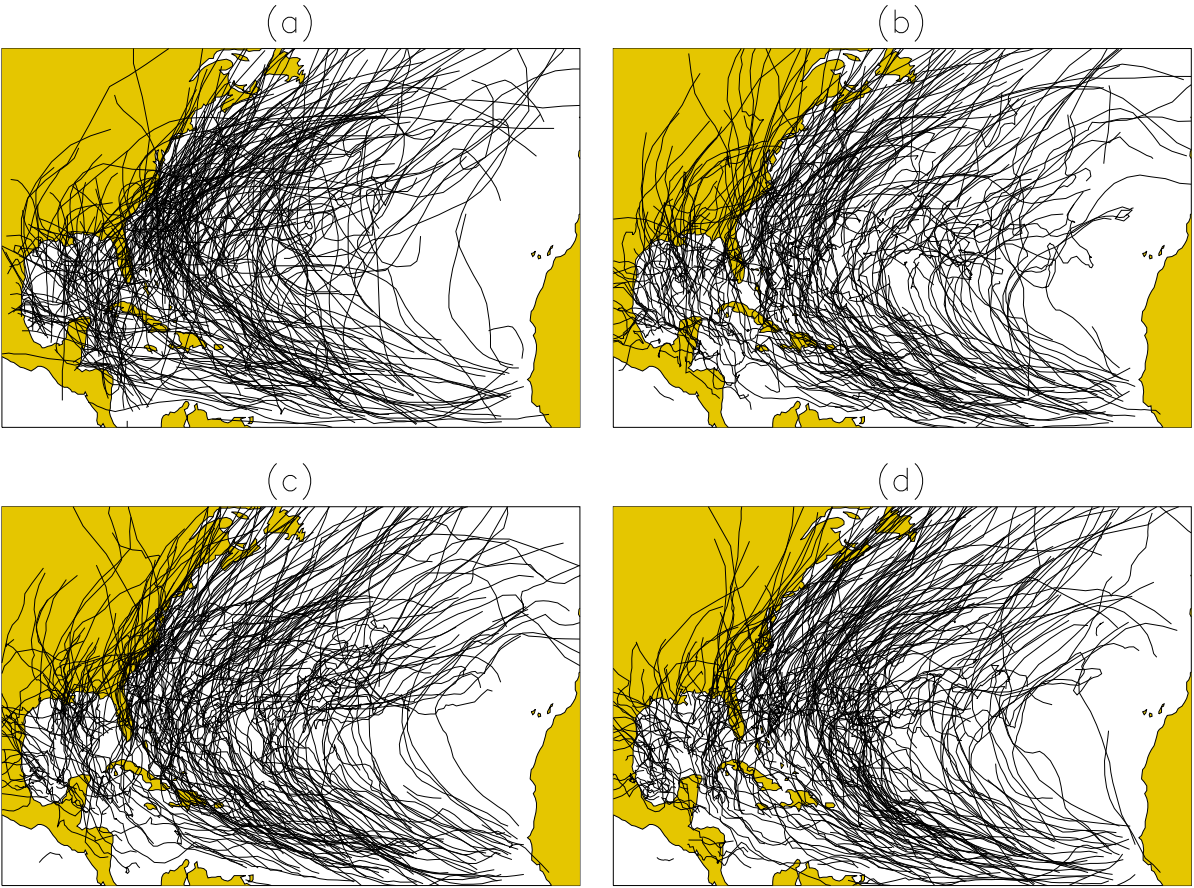}}
  \end{center}
    \caption{
Observed (panel (a)) and simulated hurricane tracks from the model described in the text.
     }
     \label{f01}
\end{figure}

\begin{figure}[!hb]
  \begin{center}
    \scalebox{0.8}{\includegraphics{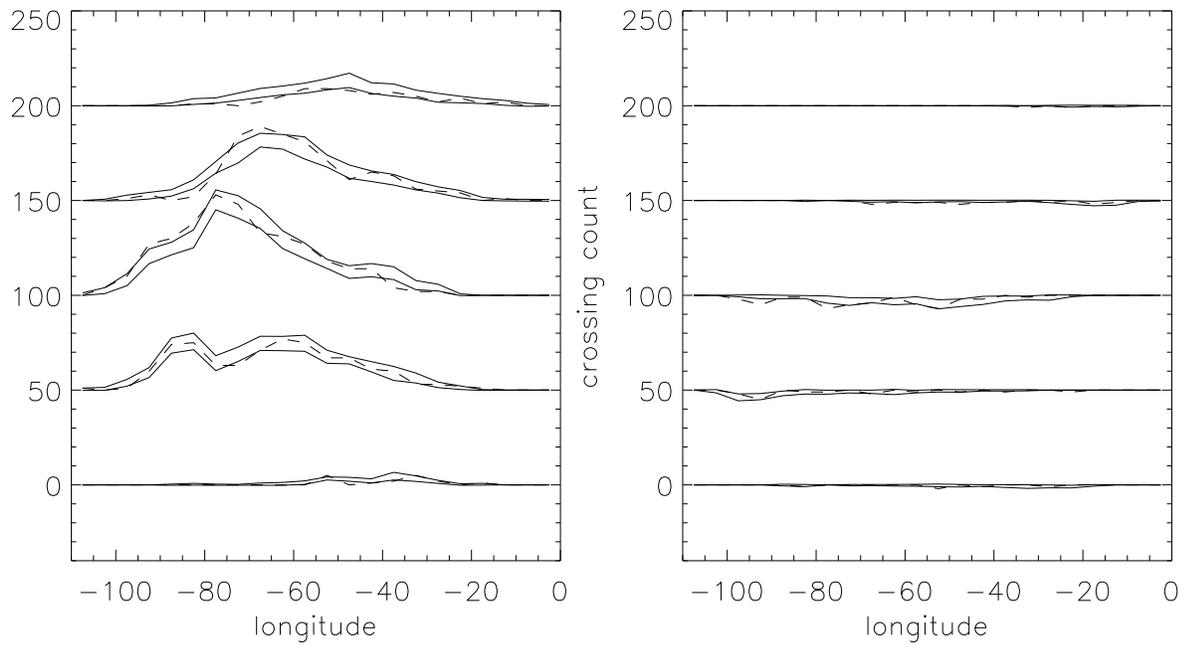}}
  \end{center}
    \caption{
  The number of tracks from observations and from simulations that cross certain lines
  of latitude (equally spaced from 10N to 50N, from bottom to top), in a northward
  direction (left panel) and in a southward direction (right panel).
     }
     \label{f02}
\end{figure}

\begin{figure}[!hb]
  \begin{center}
    \scalebox{0.8}{\includegraphics{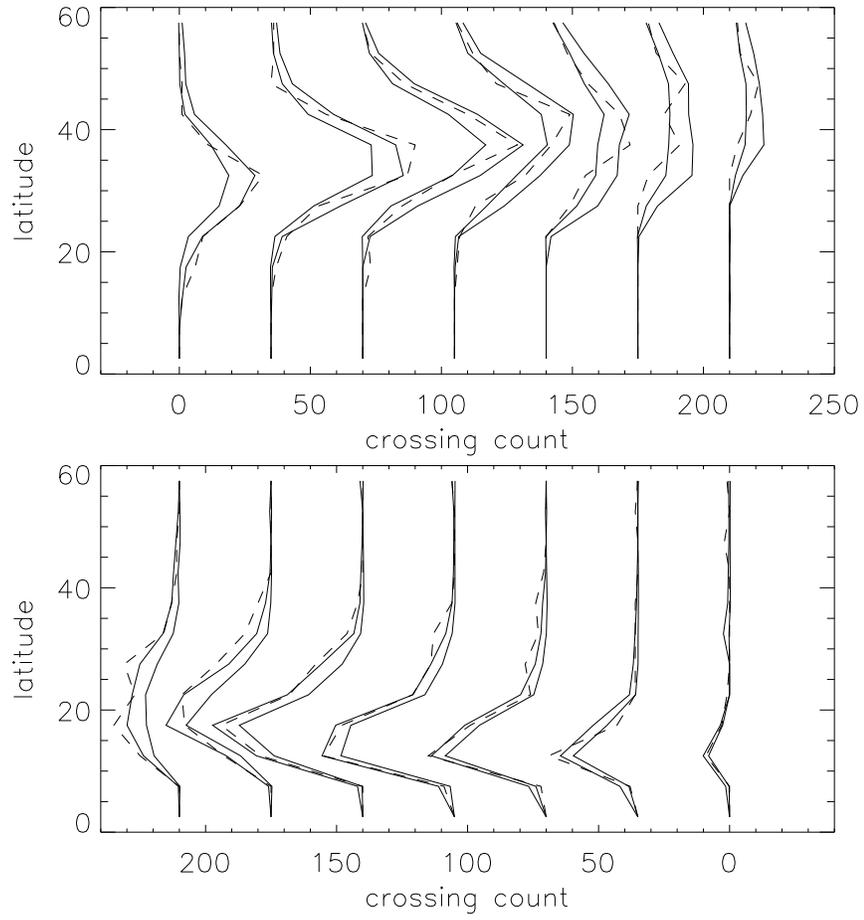}}
  \end{center}
    \caption{
  The number of tracks from observations and from simulations that cross certain lines
  of longitude (equally spaced from 80W to 20W, from left to right), in a eastward
  direction (panel (a)) and in a westward direction (panel (b)).
     }
     \label{f03}
\end{figure}

\begin{figure}[!hb]
  \begin{center}
    \scalebox{0.8}{\includegraphics{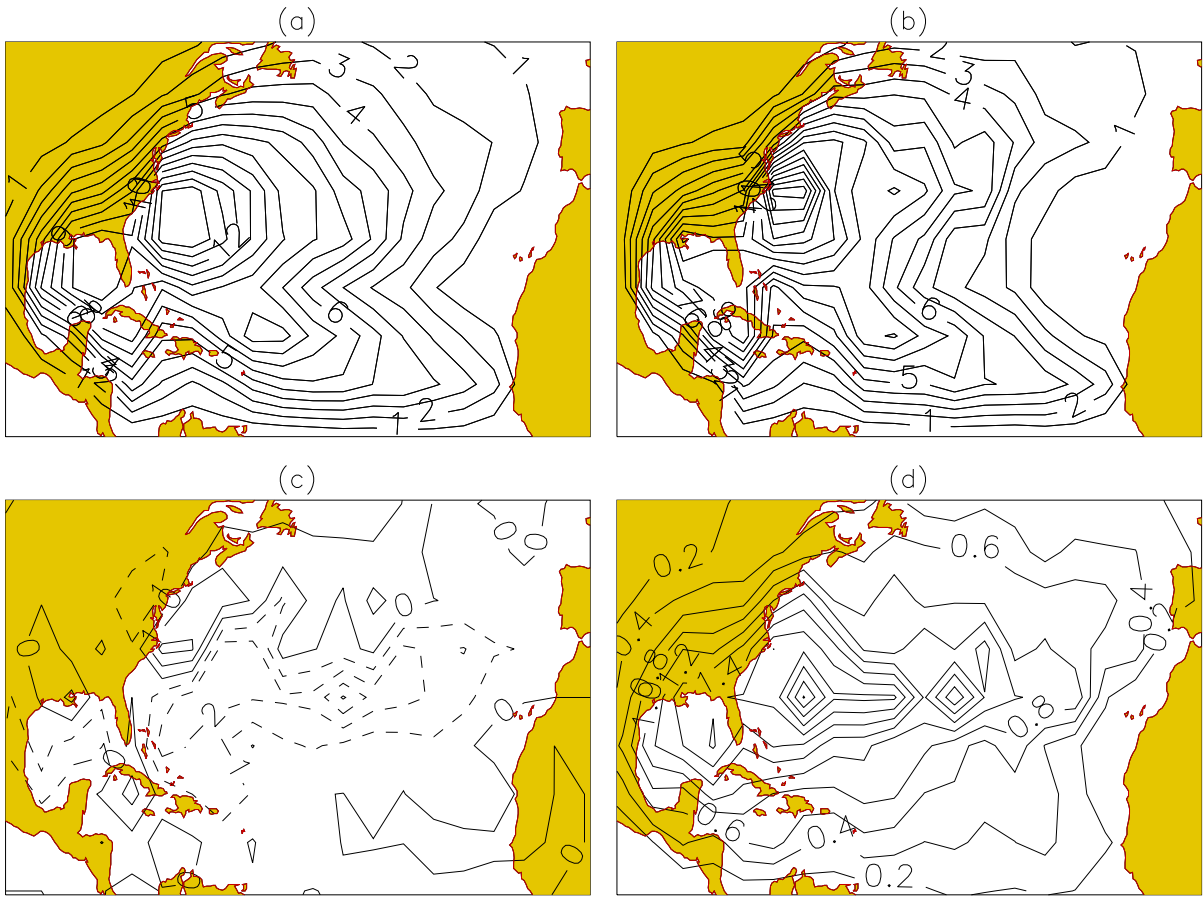}}
  \end{center}
    \caption{
  Track densities for model and observations.
  Panel (a) shows the track density for the model, averaged over 20 realisations
  of 524 storms. Panel (b) shows the track density for observations, for 524 storms.
  Panel (c) shows the difference of these densities, and panel (d) shows the
  standard deviation of the density from the model (across the 20 simulations of 524 storms).
     }
     \label{f04}
\end{figure}

\begin{figure}[!hb]
  \begin{center}
    \scalebox{0.9}{\includegraphics{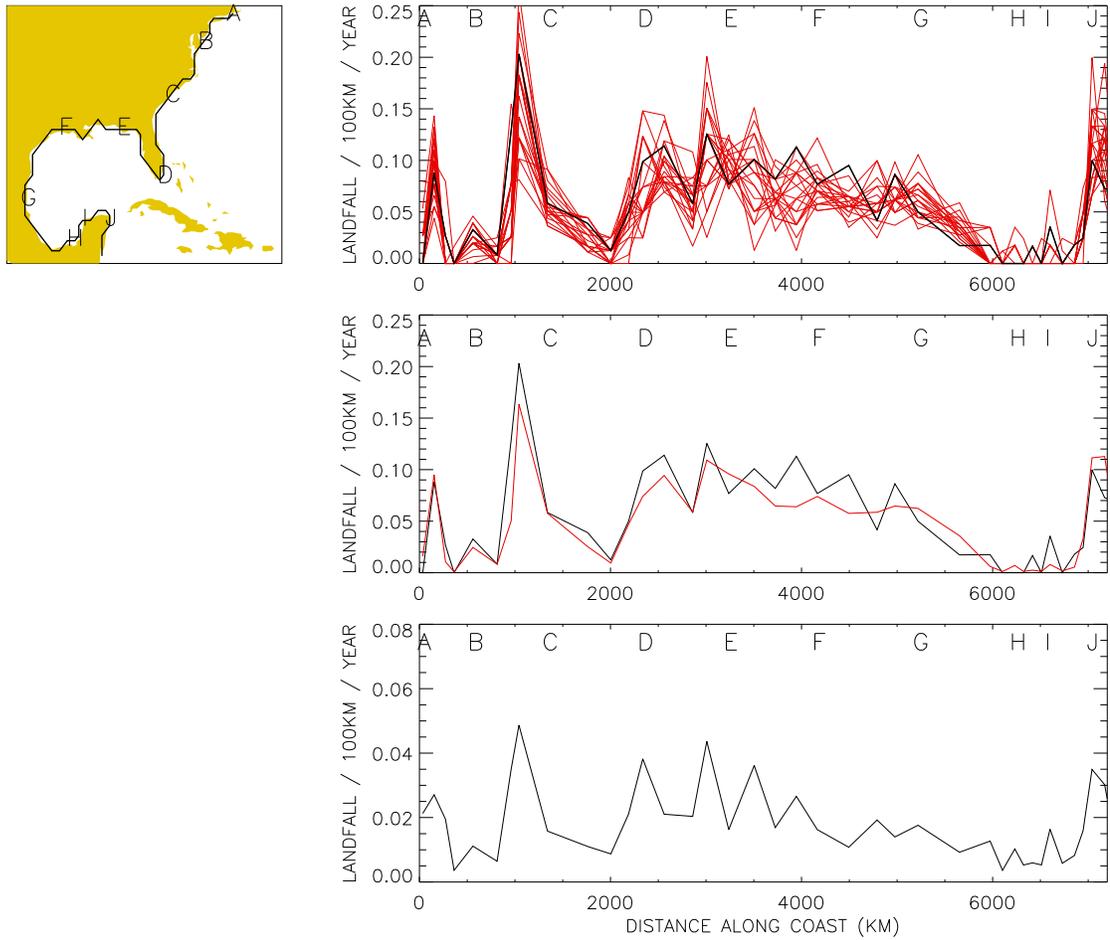}}
  \end{center}
    \caption{
The left hand panel shows a model for the coastline of North and Central America, consisting
of 39 segments. The right hand panels show various diagnostics for the number of hurricanes
crossing each of these segments in the observations and in the model.
The top panel shows 54 years of observations (black line) and 20 realisations of 54 years from the model (red lines).
The middle panel shows the observations and the mean of the model realisations.
The lower panel shows the standard deviation of the model realisations.
     }
     \label{f05}
\end{figure}

\begin{figure}[!hb]
  \begin{center}
    \scalebox{0.9}{\includegraphics{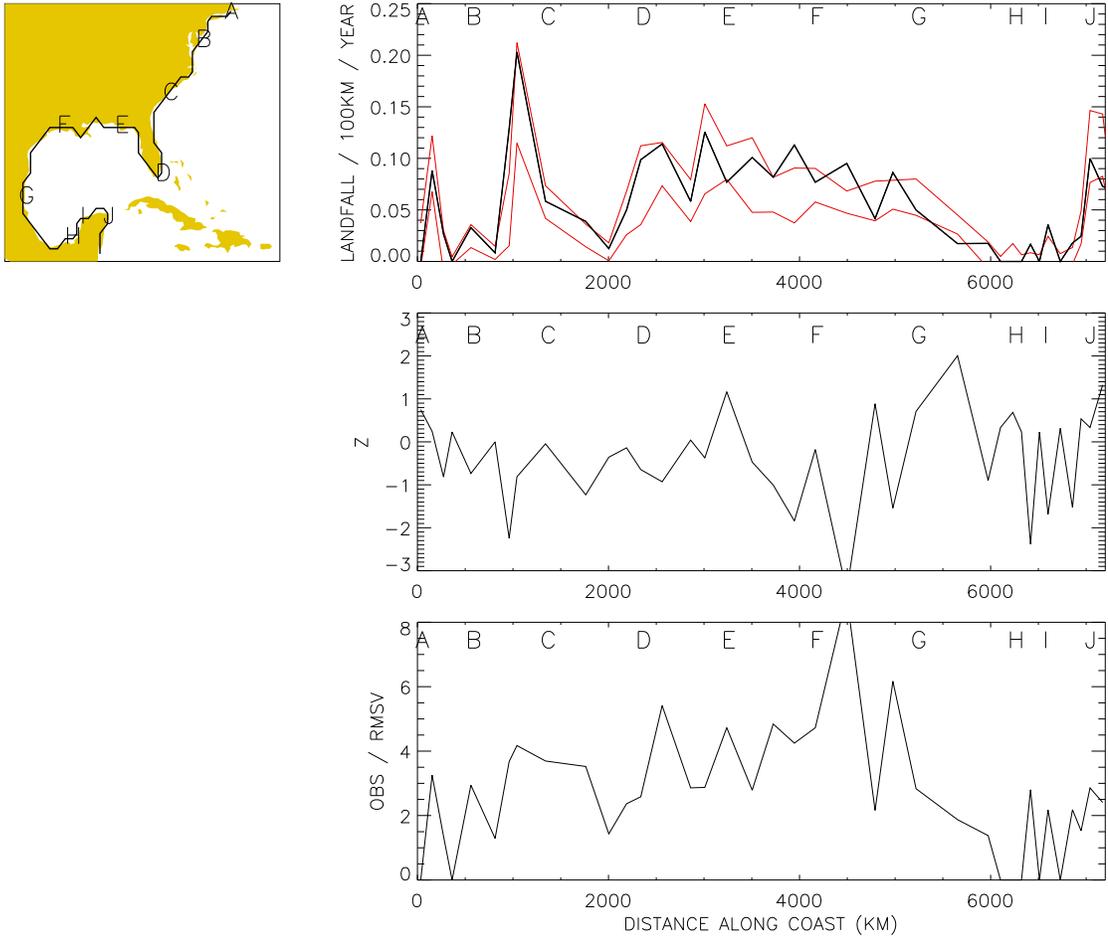}}
  \end{center}
    \caption{
As for figure~\ref{f05},but now panel (a) shows the observations, with lines for plus and minus one standard
deviation from the model, panel (b) shows the difference between model and observations normalised using the standard deviation from the
model and panel (c) shows the observations divided by the standard deviation from the model.
     }
     \label{f06}
\end{figure}

\begin{itemize}
    \item In figure~\ref{f00} we show locations of hurricane lysis from observations and the model (to be compared with
figure 2 in~\citet{hallj05e}). The results seem neither better nor worse than before.

    \item In figure~\ref{f01} we show simulated tracks (to be compared with figure 3).
The tracks look reasonable, but again it is very hard to say, by eye, whether the simulations have improved.

    \item In figure~\ref{f02} we show the rates of storms crossing lines of latitude (to be compared with
figure 4). Again, there is little difference, except perhaps a small improvement in the northward
crossing rates at the highest latitudes.

    \item In figure~\ref{f03} we show the rates of storms crossing lines of longitude (to be compared with
figure 5). The two models are not distinguishable in most places, but the new model is perhaps
slightly worse for easterly crossing in the furthest west and slightly better for westerly crossing
between 70W and 40W.

    \item In figure~\ref{f04} we see that density of storms points throughout the basin (to be compared with
6). This time, there is a definite improvement in the new model: the high density of tracks
off the coast of Cape Hatteras that is seen in the observations is now well simulated by the model.
We can perhaps attribute this to
the new simulation method using innovations which are \emph{mostly} smaller, and hence there being
a greater number of simulated storms that move along tracks similar to the mean track, and hence
end up in this particular area.

    \item In figure~\ref{f05} and figure~\ref{f06} we see various diagnostics related to landfalling rates in
the model and observations (to be compared with figures 7 and 8). Perhaps the most useful of these
diagnostics is the middle panel of figure~\ref{f06}, which shows very clearly where the model
and observations are in agreement, and where not. The results are mixed: between G and H the model
has improved, while between F and G is has got worse.

\end{itemize}

Overall, it is not possible to say whether the new model is better or worse than the previous model.
There has been an improvement in the most obvious flaw in the old model, but in other places the model
has deteriorated. That switching from normal to non-normal
innovations causes improvements in some regions, but reduces the quality of the model
in others, suggests that the behaviour of the residuals varies in space, and that modelling them as
identically distributed in space is not entirely correct.

\section{Conclusions}

We have described the latest version of our model for the simulation of hurricane tracks.
The only change from the previous version is in the modelling of the residuals: previously
we had modelled the residuals as normal, even though we knew that that wasn't entirely correct.
Now we have corrected this, and are modelling the residuals using resampling in order to capture
the correct distribution.

The results of this experiment are interesting: we have clearly improved the performance of the model
in the region where we felt the performance was most deficient, which is the model's ability to simulate
the high density of tracks just off the coast of Cape Hatteras. However, at the same time, the model's
performance has deteriorated elsewhere. This seems to imply that it would be worth increasing the complexity
of the model to allow the distribution of the residuals to vary in space, according to whatever variations
are present in the observations. This is an obvious priority for future work.

\bibliography{tim9}

%%%%%%%%%%%%%%%%%%%%%%%%%%%%%%%%%%%%%%%%%%%%%%%%%%%%%%%%%%%%%%%%%%%%%%%%%%%%%%%%%%%%%%%%%

\newpage
%%%%%%%%%%%%%%%%%%%%%%%%%%%%%%%%%%%%%%%%%%%%%%%%%%%%%%%%%%%%%%%%%%%%%%%%%%%%%%%%%%%%%%%%%

\newpage
%%%%%%%%%%%%%%%%%%%%%%%%%%%%%%%%%%%%%%%%%%%%%%%%%%%%%%%%%%%%%%%%%%%%%%%%%%%%%%%%%%%%%%%%%

\newpage
%%%%%%%%%%%%%%%%%%%%%%%%%%%%%%%%%%%%%%%%%%%%%%%%%%%%%%%%%%%%%%%%%%%%%%%%%%%%%%%%%%%%%%%%%

\newpage
%%%%%%%%%%%%%%%%%%%%%%%%%%%%%%%%%%%%%%%%%%%%%%%%%%%%%%%%%%%%%%%%%%%%%%%%%%%%%%%%%%%%%%%%%

\newpage
%%%%%%%%%%%%%%%%%%%%%%%%%%%%%%%%%%%%%%%%%%%%%%%%%%%%%%%%%%%%%%%%%%%%%%%%%%%%%%%%%%%%%%%%%

\newpage
%%%%%%%%%%%%%%%%%%%%%%%%%%%%%%%%%%%%%%%%%%%%%%%%%%%%%%%%%%%%%%%%%%%%%%%%%%%%%%%%%%%%%%%%%

\newpage

\end{document}